\documentclass{article}
\usepackage{amsmath}
\usepackage{amsmath}
\usepackage{graphicx}
\usepackage{float}
\usepackage[top=0.75in, bottom=0.75in, left=1.0in, right=1.0in]{geometry}

\input{tcilatex}

\begin{document}

\title{New Differential Formulae Related to Hermite Polynomials and their
Applications in Quantum Optics}
\author{Sun Yun$^{1}$, Wang Dong$^{1}$, Wu Jian-guang$^{1}$, and Tang Xu-bing%
$^{1,2\dag }$ \\
$^{1}$School of Mathematics \& Physics Science and Engineering, \\
Anhui University of Technology, Ma'anshan 243032, China\\
$^{2}$Institute of Intelligent Machines, Chinese Academy of Sciences,\\
Hefei 230031, China\\
$^{\dag }$ttxxbb@ahut.edu.cn}
\maketitle

\begin{abstract}
In this work, based on quantum operator Hermite polynomials and Weyl's
mapping rule, we find a generation function of the two-variable Hermite
polynomials. And then, noting that the Weyl ordering is invariant under the
similar transformations, we obtain another generalized differential
expression related to the Hermite polynomials. Those identites can be
applied to investigate the nonclasscial properties of quantum optical fields.
\end{abstract}



\section{Introduction}

As the "language" of quantum mechanics, Dirac's bra-kets have come to
represent a quantum world of abstract ideas and universal concepts and can
get a far better understanding of quantum mechanics. Because conceptions in
quantum mechanics quite differ from those in classical mechanics, it is
inevitable that quantum mechanics must have its own mathematical symbols
which are endowed with special physical meaning. For instance, in math \cite%
{Hochstadt_ie_1973}, an inhomogeneous Fredholm equation (FE) of the first
kind is written as $g\left( t\right) =\int_{-a}^{b}k\left( t,s\right)
f\left( s\right) ds,$ $K\left( t,s\right) $ is the continuous kernel
function. In quantum mechanics \cite{fhy_jopb_2003,fhy_ctp_2008}, we
introduced an operator Fredholm equation defined as $G\left( a,a^{\dag
}\right) =\int_{-a}^{b}K\left( a,a^{\dag },q\right) F\left( q\right) dq,$ in
which the kernel function $K\left( a,a^{\dag },q\right) $ is a quantum
operator, $q$ is a real variable, $a$ and $a^{\dag }$ denote the
annihilation and creation operator of a quantized radiation field. As is
well known, integrations over the operators of type $\left \vert
{}\right
\rangle \left \langle {}\right \vert $ cannot be directly
performed by Newton--Leibniz integration rule.

In Ref. \cite{fhy_ap_2006}, Fan proposed the technique of integration within
an ordered product (IWOP) of operators which enables Newton--Leibniz
integration rules directly working for Dirac's ket--bra operators with
continuum variables. The technique of IWOP shows that the operator Fredholm
equation (OFE) can directly perform integration if $K\left( a,a^{\dag
},q\right) $ is an ordered product operator. An example of taking $K\left(
a,a^{\dag },q\right) =\colon \exp \left( q-Q\right) \colon ,$ we have 
\begin{equation}
\int_{-\infty }^{\infty }\frac{dq}{\sqrt{\pi }}\colon \exp \left[ -\left(
q-Q\right) ^{2}\right] \colon f\left( q\right) =\colon G\left( Q\right)
\colon ,\text{ \  \  \  \ }Q=\frac{a+a^{\dag }}{\sqrt{2}},  \label{ope}
\end{equation}%
where $\colon \exp \left[ -\left( q-Q\right) ^{2}\right] \colon $ is the
integral kernel, $a^{\dag }$ commutes with $a$ within the normally ordered
symbol "$\colon \colon $". Noting that $\frac{1}{\sqrt{\pi }}\colon \exp %
\left[ -\left( q-Q\right) ^{2}\right] \colon =\left \vert q\right \rangle
\left \langle q\right \vert $ and the completeness relation $\int_{-\infty
}^{\infty }dq\left \vert q\right \rangle \left \langle q\right \vert =1,$ we
can see%
\begin{equation}
\int_{-\infty }^{\infty }\frac{dq}{\sqrt{\pi }}\colon \exp \left[ -\left(
q-Q\right) ^{2}\right] \colon f\left( q\right) =\int_{-\infty }^{\infty
}dq\left \vert q\right \rangle \left \langle q\right \vert f\left( q\right)
=\int_{-\infty }^{\infty }dq\left \vert q\right \rangle \left \langle
q\right \vert f\left( Q\right) =f\left( Q\right) ,  \label{exp}
\end{equation}%
where $\left \vert q\right \rangle =\pi ^{-1/4}\exp \left( -q^{2}/2+\sqrt{2}%
qa^{\dag }-a^{\dag 2}/2\right) \left \vert 0\right \rangle $ denotes the
coordinate representation with its eigenfunction $Q\left \vert
q\right
\rangle =q\left \vert q\right \rangle ,$ and $\colon G\left(
Q\right) \colon $ is the normally ordered expansion of $f\left( Q\right) .$
When $f\left( q\right) =H_{n}\left( q\right) ,$ we have%
\begin{equation}
\int_{-\infty }^{\infty }\frac{dq}{\sqrt{\pi }}\colon \exp \left[ -\left(
q-Q\right) ^{2}\right] \colon H_{n}\left( q\right) =2^{n}\colon Q^{n}\colon ,
\label{hermite}
\end{equation}%
where $2^{n}\colon Q^{n}\colon $ is the normally ordered form of the
operator Hermite polynomials $H_{n}\left( Q\right) .$ $H_{n}\left( q\right) $
is the single-variable Hermite polynomials with its generation function 
\begin{equation}
\sum_{n=0}^{\infty }\frac{q^{n}}{n!}H_{n}\left( t\right) =\exp \left(
2tq-q^{2}\right)  \label{gen1}
\end{equation}%
and $H_{n}\left( q\right) $ spans an orthonormal and complete function
space, namely $\int_{-\infty }^{\infty }\frac{dq}{\sqrt{\pi }}%
e^{-q^{2}}H_{n}\left( q\right) H_{m}\left( q\right) =2^{n}n!\delta _{nm}.$
Noting the annihilation operator $a=$ $\left( Q+iP\right) /\sqrt{2},$ and $%
\left \langle q\right \vert a^{\dag }=2^{-1/2}\left( q-\frac{d}{dq}\right)
\left \langle q\right \vert ,$ we can derive the matrix element of $%
H_{n}\left( Q\right) $ in the coordinate representation and the vacuum state 
\begin{equation*}
\left \langle q\right \vert H_{n}\left( Q\right) \left \vert 0\right \rangle
=2^{n}\left \langle q\right \vert \colon Q^{n}\colon \left \vert 0\right
\rangle ,
\end{equation*}%
and then it follows%
\begin{eqnarray}
H_{n}\left( q\right) &=&2^{n/2}e^{q^{2}/2}\left \langle q\right \vert \colon
\left( a^{\dag }+a\right) ^{n}\colon \left \vert 0\right \rangle
=2^{n/2}e^{q^{2}/2}\left \langle q\right \vert a^{\dag n}\left \vert 0\right
\rangle =e^{q^{2}/2}\left( q-\frac{d}{dq}\right) ^{n}e^{-q^{2}/2}  \notag \\
&=&\left( -1\right) ^{n}e^{q^{2}}\frac{d^{n}}{dq^{n}}e^{-q^{2}},
\label{differential}
\end{eqnarray}%
which is just the differential expression of single-variable Hermite
polynomials. The above derivation shows that Dirac's symbol and its own
arithmetic rule can promote the development of the basic quantum theory.

In Ref.\cite{erdelyi_1953_bmp}, author introduced a two-variable Hermite
polynomials in complex space 
\begin{equation}
H_{m,n}\left( \alpha ,\alpha ^{\ast }\right) =\sum_{l=0}^{\min \left(
m,n\right) }\frac{m!n!}{l!\left( m-l\right) !\left( n-l\right) !}\left(
-1\right) ^{l}\alpha ^{m-l}\alpha ^{\ast n-l},\text{ \ }\alpha =q+ip,
\label{hermite-1}
\end{equation}%
whose generating function is%
\begin{equation}
\sum_{m,n=0}^{\infty }\frac{t^{m}\tau ^{n}}{m!n!}H_{m,n}\left( \alpha
,\alpha ^{\ast }\right) =\exp \left( -t\tau +t\alpha +\tau \alpha ^{\ast
}\right) .  \label{gen2}
\end{equation}%
Two-variable Hermite polynomials can be applied in many fields of physics.
For instance, $H_{m,n}\left( \alpha ,\alpha ^{\ast }\right) $ is proved to
be the eigenmode of the complex fractional Fourier transform \cite%
{Namias_jimia_1980,Mendlovic_josaa_1993,Bernado_oc_1994}, so it may be
observed in the light propagation in graded index (GRIN) medium, and this
eigenmode is also the mechanism for two-dimensional Talbot effect
demonstrated in GRIN medium \cite{fhy_ol_2004}.

Due to there exists some similarities between the generation function of
single-variable Hermite polynomials $H_{n}\left( x\right) $ and that of
two-variable complex Hermite polynomials $H_{m,n}\left( \alpha ,\alpha
^{\ast }\right) ,$ it is interesting to see that the differential expression
of $H_{m,n}\left( \alpha ,\alpha ^{\ast }\right) $ is similar to that of $%
H_{n}\left( x\right) .$ In this work, by virtue of Weyl's mapping rule and
quantum operator Hermite polynomials, we obtain the differential expression
of $H_{m,n}\left( \alpha ,\alpha ^{\ast }\right) $ and another generalized
forms. So our work is arranged as follows. In Sec. 2, we briefly introduce
Weyl correspondence rule, which is related to the Weyl ordering and the
technique of integration within Weyl ordered product (IWWP) of operators. We
reveal that quantum correspondence operator of classical function $f\left(
p,q\right) $ can directly be obtianed by replacing $q$ and $p$ in $f\left(
p,q\right) $ by $Q$ and $P$ with the function form invariant. In Sec. 3 we
introduce the two-variable Hermite function. By formulizing the Weyl
correspondence, we derive a differential expression of $H_{m,n}\left( \alpha
,\alpha ^{\ast }\right) .$ Noting that the Weyl ordering is invariant under
the similar transformations, the generalized differential expressions are
obtained in Sec. 4. Enlightened by those new identities, in Sec. V we
investigate the nonclassical properties of an excited squeezed vacuum state,
which can be generated in quantum systems with restricted dimensions.

\section{Weyl Ordering and Weyl Correspondence Rule}

As is well known, the Weyl correspondence rule \cite%
{Weyl_zp_1927,Weyl_tgqm_2007}, i.e.%
\begin{equation}
F\left( P,Q\right) =\int \int dqdpf\left( p,q\right) \Delta \left(
q,p\right) ,  \label{weyl1}
\end{equation}%
is a recipe for quantizing a classical function $f\left( p,q\right) $
defined in classical phase space as a quantum correspond operator $F\left(
P,Q\right) $. $\Delta \left( q,p\right) $ is the Wigner operator difined as 
\begin{equation}
\Delta \left( p,q\right) =\int_{-\infty }^{\infty }\frac{du}{2\pi }%
e^{ipu}\left \vert q+\frac{u}{2}\right \rangle \left \langle q-\frac{u}{2}%
\right \vert .  \label{wig}
\end{equation}%
When $f\left( p,q\right) =q^{m}p^{r},$ from Eq. (\ref{weyl1}) its quantum
corresponding operators is 
\begin{equation}
q^{m}p^{r}\rightarrow \left( \frac{1}{2}\right) ^{m}\sum_{l=0}^{m}\frac{m!}{%
l!\left( m-l\right) !}Q^{m-l}P^{r}Q^{l}=\left( \frac{1}{2}\right)
^{m}\sum_{l=0}^{m}\frac{m!}{l!\left( m-l\right) !}%
\begin{array}{c}
\dot{\vdots}%
\end{array}%
Q^{m-l}P^{r}Q^{l}%
\begin{array}{c}
\dot{\vdots}%
\end{array}%
=%
\begin{array}{c}
\dot{\vdots}%
\end{array}%
Q^{m}P^{r}%
\begin{array}{c}
\dot{\vdots}%
\end{array}%
,  \label{4}
\end{equation}%
in which $%
\begin{array}{c}
\dot{\vdots}%
\end{array}%
\begin{array}{c}
\dot{\vdots}%
\end{array}%
$\ denotes the Weyl ordering symbol, and Eq. (\ref{4})\ tell us that the
Weyl ordered correspond operator of classcial function $f\left( p,q\right) $
is obtained by just replacing $p,q$ in $f\left( p,q\right) $ by $P,Q\ $\
with the function form invariant. As one of the definite operator orderings
(such as normal ordering, anti-normal ordering and Weyl ordering), Weyle
ordering is a useful one and within Weyl ordering symbol $%
\begin{array}{c}
\dot{\vdots}%
\end{array}%
\begin{array}{c}
\dot{\vdots}%
\end{array}%
$ the Bose operators are permutable. Enlighted by the technique of IWOP, 
\cite{fhy_jpa_1992} proposed the technique of integration within Weyl
ordered product (IWWP) of operators. From (\ref{wig}) we obtain the Weyl
ordering form of the Wigner opertor%
\begin{equation}
\Delta \left( p,q\right) =%
\begin{array}{c}
\dot{\vdots}%
\end{array}%
\delta \left( p-P\right) \delta \left( q-Q\right) 
\begin{array}{c}
\dot{\vdots}%
\end{array}%
.  \label{wig2}
\end{equation}%
Therefore one can easily obtain quantum correspond operator of classical
function $h\left( p,q\right) $ by replacing $q\rightarrow Q,$ $p\rightarrow
P $, i.e.%
\begin{equation}
F\left( P,Q\right) =%
\begin{array}{c}
\dot{\vdots}%
\end{array}%
h\left( P,Q\right) 
\begin{array}{c}
\dot{\vdots}%
\end{array}%
=\int \int_{-\infty }^{\infty }dpdqh\left( p,q\right) \Delta \left(
p,q\right) .  \label{weyl2}
\end{equation}%
Eq. (\ref{ope}) can tell us that the Weyl correspondence rule is also an
OFE. Noting that 
\begin{eqnarray}
\text{Tr}\left[ \Delta \left( p_{1},q_{1}\right) \Delta \left(
p_{2},q_{2}\right) \right] &=&\int \frac{d^{2}z}{\pi }\left \langle z\right
\vert \left[ \Delta \left( p_{1},q_{1}\right) \Delta \left(
p_{2},q_{2}\right) \right] \left \vert z\right \rangle  \notag \\
&=&\frac{1}{2\pi }\delta \left( q_{1}-q_{2}\right) \delta \left(
p_{1}-p_{2}\right) ,  \label{doubw}
\end{eqnarray}%
it then follows that the reciprocal relation of the Weyl correspondence rule
is 
\begin{equation}
2\pi Tr\left[ F\left( P,Q\right) \Delta \left( q,p\right) \right] =2\pi 
\text{Tr}\left[ \int \int dq_{1}dp_{1}h\left( p_{1},q_{1}\right) \Delta
\left( p_{1},q_{1}\right) \Delta \left( p,q\right) \right] =h\left(
p,q\right) .  \label{reciprocal}
\end{equation}

In many cases, taking $\alpha =\left( q+ip\right) /\sqrt{2}$, the Wigner
operator in (\ref{wig}) can be rewritten as%
\begin{eqnarray}
\Delta \left( p,q\right) &\rightarrow &\Delta \left( \alpha ,\alpha ^{\ast
}\right) =\int \frac{d^{2}z}{\pi ^{2}}\left \vert \alpha +z\right \rangle
\left \langle \alpha -z\right \vert e^{\alpha z^{\ast }-\alpha ^{\ast }z} 
\notag \\
&=&\frac{1}{\pi }\colon \exp \left[ -2\left( a^{\dagger }-\alpha ^{\ast
}\right) \left( a-\alpha \right) \right] \colon =\frac{1}{2}%
\begin{array}{c}
\dot{\vdots}%
\end{array}%
\delta \left( a^{\dagger }-\alpha ^{\ast }\right) \delta \left( a-\alpha
\right) 
\begin{array}{c}
\dot{\vdots}%
\end{array}%
.  \label{wig3}
\end{eqnarray}%
It then follows that the Weyl correspondence formula in Eq. (\ref{weyl2})
can be recast to 
\begin{equation}
G\left( a,a^{\dagger }\right) =%
\begin{array}{c}
\dot{\vdots}%
\end{array}%
f\left( a,a^{\dagger }\right) 
\begin{array}{c}
\dot{\vdots}%
\end{array}%
=2\int \int dqdpf\left( \alpha ,\alpha ^{\ast }\right) \Delta \left( \alpha
,\alpha ^{\ast }\right) ,  \label{weyl3}
\end{equation}%
with its reciprocal relation%
\begin{equation}
2\pi Tr\left[ G\left( a,a^{\dagger }\right) \Delta \left( \alpha ,\alpha
^{\ast }\right) \right] =2\pi Tr\left[ 
\begin{array}{c}
\dot{\vdots}%
\end{array}%
f\left( a,a^{\dagger }\right) 
\begin{array}{c}
\dot{\vdots}%
\end{array}%
\Delta \left( \alpha ,\alpha ^{\ast }\right) \right] =f\left( \alpha ,\alpha
^{\ast }\right) .  \label{rr}
\end{equation}%
Especially, when $G\left( a,a^{\dagger }\right) =\rho \left( a,a^{\dagger
}\right) $ describes a density of states for an interesting quantum system,
from the above reciprocal relation we can see 
\begin{equation}
2\pi Tr\left[ \rho \left( a,a^{\dagger }\right) \Delta \left( \alpha ,\alpha
^{\ast }\right) \right] =W\left( \alpha ,\alpha ^{\ast }\right) ,  \label{ww}
\end{equation}%
$W\left( \alpha ,\alpha ^{\ast }\right) $ denotes a quasi-probability
distribution Wigner function.\ Thus Eq. (\ref{ww}) can also be called the
Wigner-Weyl correspondence rule.

\section{New Differential Expression of Two-variable Hermite Polynomials $%
H_{m,n}\left( \protect \alpha ,\protect \alpha ^{\ast }\right) $}

In quantum theory, $H_{m,n}\left( \alpha ,\alpha ^{\ast }\right) $ is the
generalized Bargmann representation of the two-mode Fock state in the
bipartite entangled state representation \cite{fhy_pla_2002}, i.e.%
\begin{equation*}
\left \vert m,n\right \rangle =\frac{a^{\dagger m}b^{\dagger m}}{\sqrt{m!n!}}%
\left \vert 0,0\right \rangle \rightarrow \frac{1}{\sqrt{m!n!}}H_{m,n}\left(
\xi ,\xi ^{\ast }\right) e^{-\frac{\left \vert \xi \right \vert ^{2}}{2}},
\end{equation*}%
and it spans an orthonormal and complete function space,%
\begin{equation}
2\int \int \frac{d^{2}\xi }{\pi }e^{-2\left \vert \xi \right \vert
^{2}}H_{m,n}\left( \sqrt{2}\xi ,\sqrt{2}\xi ^{\ast }\right) H_{m^{\prime
},n^{\prime }}^{\ast }\left( \sqrt{2}\xi ,\sqrt{2}\xi ^{\ast }\right) =\sqrt{%
m!n!m^{\prime }!n^{\prime }!}\delta _{m,m^{\prime }}\delta _{n,n^{\prime }}.
\label{orthonor}
\end{equation}%
Eq.(\ref{orthonor}) indicates any one function $f\left( \alpha ,\alpha
^{\ast }\right) $ can be expanded by those orthogonal basis,%
\begin{equation}
f\left( \alpha ,\alpha ^{\ast }\right) =\sum_{m,n=0}^{\infty
}C_{m,n}H_{m,n}^{\ast }\left( \sqrt{2}\alpha ,\sqrt{2}\alpha ^{\ast }\right)
.  \label{expand}
\end{equation}%
where $C_{m,n}$ is a constant to be determined by follow derivation. From
its generation function shown in (\ref{gen2}), we can also expand the
normally ordered form of Wigner operator in Eq. (\ref{wig3}) as%
\begin{equation}
\Delta \left( \alpha ,\alpha ^{\ast }\right) =\frac{1}{\pi }\colon \exp %
\left[ -2\left( a^{\dagger }-\alpha ^{\ast }\right) \left( a-\alpha \right) %
\right] \colon =\frac{1}{\pi }e^{-2\left \vert \alpha \right \vert
^{2}}\colon \sum_{m,n=0}^{\infty }\frac{\sqrt{2^{m+n}}a^{\dag m}a^{n}}{m!n!}%
H_{m,n}\left( \sqrt{2}\alpha ,\sqrt{2}\alpha ^{\ast }\right) \colon .
\label{w_ordering}
\end{equation}%
Substituting (\ref{expand}) and (\ref{w_ordering}) into (\ref{weyl3}), we
have%
\begin{eqnarray}
G\left( a^{\dag },a\right) &=&2\int \frac{d^{2}\alpha }{\pi }e^{-2\left
\vert \alpha \right \vert ^{2}}\colon \sum_{m,n=0}^{\infty }\frac{\sqrt{%
2^{m+n}}a^{\dag m}a^{n}}{m!n!}H_{m,n}\left( \sqrt{2}\alpha ,\sqrt{2}\alpha
^{\ast }\right) \colon \sum_{m^{\prime },n^{\prime }=0}^{\infty
}C_{m^{\prime },n^{\prime }}H_{m^{\prime },n^{\prime }}^{\ast }\left( \sqrt{2%
}\alpha ,\sqrt{2}\alpha ^{\ast }\right)  \notag \\
&=&2\colon \sum_{m,n=0}^{\infty }\sum_{m^{\prime },n^{\prime }=0}^{\infty
}C_{m^{\prime },n^{\prime }}\frac{\sqrt{2^{m+n}}a^{\dag m}a^{n}}{m!n!}\colon
\int \frac{d^{2}\alpha }{\pi }e^{-2\left \vert \alpha \right \vert
^{2}}H_{m,n}\left( \sqrt{2}\alpha ,\sqrt{2}\alpha ^{\ast }\right)
H_{m^{\prime },n^{\prime }}^{\ast }\left( \sqrt{2}\alpha ,\sqrt{2}\alpha
^{\ast }\right)  \label{expand2} \\
&=&\colon \sum_{m,n=0}^{\infty }C_{m,n}\sqrt{2^{m+n}}a^{\dag m}a^{n}\colon
\equiv \colon F\left( a^{\dag },a\right) \colon .  \notag
\end{eqnarray}%
Taking the coherent state expectation values of (\ref{expand2}), we have%
\begin{equation*}
\left \langle \alpha \right \vert \colon F\left( a^{\dag },a\right) \colon
\left \vert \alpha \right \rangle =F\left( \alpha ^{\ast },\alpha \right)
=\left \langle \alpha \right \vert \colon \sum_{m,n=0}^{\infty }C_{m,n}\sqrt{%
2^{m+n}}a^{\dag m}a^{n}\colon \left \vert \alpha \right \rangle
=\sum_{m,n=0}^{\infty }C_{m,n}\sqrt{2^{m+n}}\left( \alpha ^{\ast }\right)
^{m}\alpha ^{n},
\end{equation*}%
and then 
\begin{equation*}
C_{m,n}=\frac{1}{m!n!\sqrt{2^{m+n}}}\left. \frac{\partial {}^{m}}{\partial
\alpha ^{\ast }{}^{m}}\frac{\partial {}^{n}}{\partial \alpha {}^{n}}F\left(
\alpha ^{\ast },\alpha \right) \right \vert _{\alpha =0}.
\end{equation*}%
Therefore, from (\ref{expand}) we obtain%
\begin{equation}
F\left( \alpha ,\alpha ^{\ast }\right) =\sum_{m,n=0}^{\infty }H_{m,n}\left( 
\sqrt{2}\alpha ,\sqrt{2}\alpha ^{\ast }\right) \frac{1}{m!n!\sqrt{2^{m+n}}}%
\left. \frac{\partial {}^{m}}{\partial \alpha ^{\ast }{}^{m}}\frac{\partial
{}^{n}}{\partial \alpha {}^{n}}F\left( \alpha ^{\ast },\alpha \right) \right
\vert _{\alpha =0}.  \label{10}
\end{equation}%
This is a new formula for deriving Weyl's classical correspondence of
normally ordered quantum operators. For example, when $G_{1}\left( a^{\dag
},a\right) =\colon F_{1}\left( a^{\dag },a\right) \colon =\colon a^{\dag
m}a^{n}\colon ,$ from Eq. (\ref{10}) we have%
\begin{equation}
g_{1}\left( \alpha ,\alpha ^{\ast }\right) =\frac{1}{\sqrt{2^{m+n}}}%
H_{m,n}\left( \sqrt{2}\alpha ,\sqrt{2}\alpha ^{\ast }\right) .
\label{correspond}
\end{equation}%
And then considering the reciprocal relation in (\ref{rr}), we can see 
\begin{equation*}
g_{1}\left( \alpha ,\alpha ^{\ast }\right) =2\pi \text{Tr}\left[ G_{1}\left(
a^{\dag },a\right) \Delta \left( \alpha ,\alpha ^{\ast }\right) \right]
=2\pi \text{Tr}\left[ a^{\dag m}a^{n}\Delta \left( \alpha ,\alpha ^{\ast
}\right) \right] =\frac{1}{\sqrt{2^{m+n}}}H_{m,n}\left( \sqrt{2}\alpha ,%
\sqrt{2}\alpha ^{\ast }\right)
\end{equation*}

On the other hand, given any operator $F\left( a,a^{\dagger }\right) ,$\ its
classical correspondence can also be obtained by utilizing the following
formular (see Appendix for details)%
\begin{equation}
f\left( \alpha ^{\ast },\alpha \right) =e^{2\left \vert \alpha \right \vert
^{2}}\int \frac{d^{2}\beta }{\pi }\left \langle -\beta \right \vert F\left(
a^{\dagger },a\right) \left \vert \beta \right \rangle e^{2\left( \alpha
\beta ^{\ast }-\alpha ^{\ast }\beta \right) }.  \label{formula}
\end{equation}%
Substituting $G_{1}\left( a^{\dag },a\right) =\colon a^{\dag m}a^{n}\colon $
into (\ref{formula}), we have%
\begin{eqnarray*}
g_{1}\left( \alpha ,\alpha ^{\ast }\right) &=&e^{2\left \vert \alpha \right
\vert ^{2}}\int \frac{d^{2}\beta }{\pi }\left \langle -\beta \right \vert
G_{1}\left( a^{\dag },a\right) \left \vert \beta \right \rangle e^{2\left(
\alpha \beta ^{\ast }-\alpha ^{\ast }\beta \right) }=e^{2\left \vert \alpha
\right \vert ^{2}}\int \frac{d^{2}\beta }{\pi }\left \langle -\beta \right
\vert \colon a^{\dag m}a^{n}\colon \left \vert \beta \right \rangle
e^{2\left( \alpha \beta ^{\ast }-\alpha ^{\ast }\beta \right) } \\
&=&e^{2\left \vert \alpha \right \vert ^{2}}\int \frac{d^{2}\beta }{\pi }%
\left( -\beta ^{\ast }\right) ^{m}\beta ^{n}e^{-2\left \vert \beta \right
\vert ^{2}+2\left( \alpha \beta ^{\ast }-\alpha ^{\ast }\beta \right) }.
\end{eqnarray*}%
Making the substitution $\sqrt{2}\beta \rightarrow \beta ,\sqrt{2}\beta
^{\ast }\rightarrow \beta ^{\ast },$ the above equation can be derived as%
\begin{eqnarray*}
g_{1}\left( \alpha ,\alpha ^{\ast }\right) &=&\frac{\left( -1\right)
^{m}e^{2\left \vert \alpha \right \vert ^{2}}}{\left( \sqrt{2}\right) ^{m+n}}%
\int \frac{d^{2}\beta }{\pi }\beta ^{\ast m}\beta ^{n}\exp \left[ -\left
\vert \beta \right \vert ^{2}+\sqrt{2}\left( \alpha \beta ^{\ast }-\alpha
^{\ast }\beta \right) \right] \\
&=&\frac{\left( -1\right) ^{m-n}e^{2\left \vert \alpha \right \vert ^{2}}}{%
2^{m+n}}\frac{\partial {}^{m}}{\partial \alpha {}^{m}}\frac{\partial {}^{n}}{%
\partial \alpha ^{\ast }{}^{n}}\int \frac{d^{2}\beta }{\pi }\exp \left[
-\left \vert \beta \right \vert ^{2}+\sqrt{2}\left( \alpha \beta ^{\ast
}-\alpha ^{\ast }\beta \right) \right] .
\end{eqnarray*}%
Utilizing the following integral formula%
\begin{equation*}
\int \frac{d^{2}\alpha }{\pi }\exp \left[ h\left \vert \alpha \right \vert
^{2}+s\alpha +\eta \alpha ^{\ast }\right] =\frac{1}{h}\exp \left[ -\frac{%
s\eta }{h}\right] ,\text{ \ Re}\left[ h\right] <0,
\end{equation*}%
we can obtain%
\begin{equation}
g_{1}\left( \alpha ,\alpha ^{\ast }\right) =\frac{\left( -1\right)
^{m-n}e^{2\left \vert \alpha \right \vert ^{2}}}{2^{m+n}}\frac{\partial
{}^{m}}{\partial \alpha {}^{m}}\frac{\partial {}^{n}}{\partial \alpha ^{\ast
}{}^{n}}\exp \left( -2\left \vert \alpha \right \vert ^{2}\right) .
\label{wigner}
\end{equation}%
\ Comparing Eq. (\ref{wigner}) and (\ref{correspond}), and setting $\sqrt{2}%
\alpha \rightarrow \alpha ,\sqrt{2}\alpha ^{\ast }\rightarrow \alpha ^{\ast
} $ we can derive%
\begin{equation}
H_{m,n}\left( \alpha ,\alpha ^{\ast }\right) =\left( -1\right)
^{m-n}e^{\left \vert \alpha \right \vert ^{2}}\frac{\partial {}^{m}}{%
\partial \alpha ^{\ast }{}^{m}}\frac{\partial {}^{n}}{\partial \alpha {}^{n}}%
\exp \left( -\left \vert \alpha \right \vert ^{2}\right) ,  \label{formular}
\end{equation}%
which is just the differential form for the generation function of
two-variable Hermite polynomials, and for the especial case of $m=n$%
\begin{equation*}
H_{m,m}\left( \alpha ,\alpha ^{\ast }\right) =\exp \left( \left \vert \alpha
\right \vert ^{2}\right) \frac{\partial {}^{2m}}{\partial \alpha ^{\ast
}{}{}^{m}\alpha ^{m}}\exp \left( -\left \vert \alpha \right \vert
^{2}\right) .
\end{equation*}

\section{Generalized Differential Expressions related to Hermite Polynomials}

In order to generate non-symmetric quantum mechanical representation, in Ref.%
\cite{fhy_1991_jpa} authors proposed a non-unitary operator $\hat{U}$,
defined as%
\begin{eqnarray}
\hat{U} &=&\frac{1}{\sqrt{\mu }}\int \frac{d^{2}z}{\pi }\left \vert \left( 
\begin{array}{cc}
\mu & \nu \\ 
\sigma & \tau%
\end{array}%
\right) \left( 
\begin{array}{c}
z \\ 
z^{\ast }%
\end{array}%
\right) \right \rangle \left \langle \left( 
\begin{array}{c}
z \\ 
z^{\ast }%
\end{array}%
\right) \right \vert  \notag \\
&=&\frac{1}{\sqrt{\mu }}\colon \exp \left[ -\frac{\nu }{2\mu }a^{\dagger
2}+\left( \frac{1}{\mu }-1\right) a^{\dagger }a+\frac{\sigma }{2\mu }a^{2}%
\right] \colon ,  \label{un1}
\end{eqnarray}%
which is a quantum operator image of the classical symplectic transformation 
$\left( z,z^{\ast }\right) \rightarrow \left( \mu z+\nu z^{\ast },\sigma
z+\tau z^{\ast }\right) $ in phase space$,$ and where $\left \vert \left( 
\begin{array}{c}
z \\ 
z^{\ast }%
\end{array}%
\right) \right \rangle =\exp \left( za^{\dagger }-z^{\ast }a\right)
\left
\vert 0\right \rangle $ denotes the coherent state \cite%
{Glauber_1963_pr}\ and $\left \vert \left( 
\begin{array}{cc}
\mu & \nu \\ 
\sigma & \tau%
\end{array}%
\right) \left( 
\begin{array}{c}
z \\ 
z^{\ast }%
\end{array}%
\right) \right \rangle =\left \vert \left( 
\begin{array}{c}
\mu z+\nu z^{\ast } \\ 
\sigma z+\tau z^{\ast }%
\end{array}%
\right) \right \rangle =\exp \left[ \left( \mu z+\nu z^{\ast }\right)
a^{\dagger }-\left( \sigma z+\tau z^{\ast }\right) a\right] \left \vert
0\right \rangle .$ The invers of $\hat{U}$\ reads as%
\begin{eqnarray}
\hat{U}^{-1} &=&\sqrt{\mu }\int \frac{d^{2}z}{\pi }\left \vert \left( 
\begin{array}{c}
z \\ 
z^{\ast }%
\end{array}%
\right) \right \rangle \left \langle \left( 
\begin{array}{cc}
\tau & -\nu \\ 
-\sigma & \mu%
\end{array}%
\right) \left( 
\begin{array}{c}
z \\ 
z^{\ast }%
\end{array}%
\right) \right \vert  \notag \\
&=&\frac{1}{\sqrt{\tau }}\colon \exp \left[ \frac{\nu }{2\tau }a^{\dagger
2}+\left( \frac{1}{\tau }-1\right) a^{\dagger }a-\frac{\sigma }{2\tau }a^{2}%
\right] \colon .  \label{un2}
\end{eqnarray}%
\ From (\ref{un1}) and (\ref{un2}), we can find $\hat{U}^{\dag }\neq \hat{U}%
^{-1}$ and $\hat{U}$\ engenders a similar transformation 
\begin{equation}
\hat{U}a\hat{U}^{-1}=\mu a+\nu a^{\dagger },\  \hat{U}a^{\dagger }\hat{U}%
^{-1}=\sigma a+\tau a^{\dagger }  \label{similar_trans}
\end{equation}%
and its invers transformation%
\begin{equation}
\hat{U}^{-1}a\hat{U}=\tau a-\nu a^{\dagger },\  \hat{U}^{-1}a^{\dagger }\hat{U%
}=\mu a^{\dagger }-\sigma a,  \label{invers}
\end{equation}%
where four complex parameters satisfies $\mu \tau -\nu \sigma =1$ for
keeping $\left[ \mu a+\nu a^{\dagger },\  \sigma a+\tau a^{\dagger }\right]
=1.$Also it is important to note that the Weyl ordering is invariant under
the similar transformations \cite{fhy_ap_2006,fhy_ctp_2003,chen_ijtp_2012},
i.e. 
\begin{equation}
\hat{U}G\left( a^{\dagger },a\right) \hat{U}^{-1}=\hat{U}%
\begin{array}{c}
\dot{\vdots}%
\end{array}%
g\left( a^{\dagger },a\right) 
\begin{array}{c}
\dot{\vdots}%
\end{array}%
\hat{U}^{-1}=%
\begin{array}{c}
\dot{\vdots}%
\end{array}%
g\left( \sigma a+\tau a^{\dagger },\mu a+\nu a^{\dagger }\right) 
\begin{array}{c}
\dot{\vdots}%
\end{array}%
.  \label{invariant}
\end{equation}

\subsection{Generalized Differential Expression Related to the Product of
Two Single-variable Hermite Polynomials}

Supposing an operator $G_{2}\left( a^{\dag },a\right) =a^{\dagger m}\hat{U}%
\left \vert 0\right \rangle \left \langle 0\right \vert \hat{U}^{-1}a^{n},$
and from Eqs. (\ref{rr}) and (\ref{invers}), we can express the classical
Weyl correspondence of $G_{2}\left( a^{\dagger },a\right) $ as 
\begin{eqnarray}
g_{2}\left( \alpha ^{\ast },\alpha \right) &=&2\pi \text{Tr}\left[
G_{2}\left( a^{\dagger },a\right) \Delta \left( \alpha ,\alpha ^{\ast
}\right) \right] =2\pi \text{Tr}\left[ a^{\dagger m}\hat{U}\left \vert
0\right \rangle \left \langle 0\right \vert \hat{U}^{-1}a^{n}\Delta \left(
\alpha ,\alpha ^{\ast }\right) \right]  \notag \\
&=&2\pi \text{Tr}\left[ \hat{U}\left( \mu a^{\dagger }-\sigma a\right)
^{m}\left \vert 0\right \rangle \left \langle 0\right \vert \left( \tau
a-\nu a^{\dagger }\right) ^{n}\hat{U}^{-1}\Delta \left( \alpha ,\alpha
^{\ast }\right) \right] .  \label{weylc}
\end{eqnarray}%
By virtue of the following formular%
\begin{equation*}
\left( fa+ga^{\dag }\right) ^{n}=\left( -i\sqrt{\frac{fg}{2}}\right)
^{n}\colon H_{n}\left( i\sqrt{\frac{f}{2g}}a+i\sqrt{\frac{g}{2f}}a^{\dag
}\right) \colon ,
\end{equation*}%
we can see%
\begin{eqnarray}
\left( \mu a^{\dagger }-\sigma a\right) ^{m} &=&\left( \sqrt{\frac{\sigma
\mu }{2}}\right) ^{m}\colon H_{m}\left( -\sqrt{\frac{\sigma }{2\mu }}a-\sqrt{%
\frac{\mu }{2\sigma }}a^{\dag }\right) \colon  \label{32} \\
\left( \tau a-\nu a^{\dagger }\right) ^{n} &=&\left( \sqrt{\frac{\nu \tau }{2%
}}\right) ^{n}\colon H_{n}\left( -\sqrt{\frac{\tau }{2\nu }}a-\sqrt{\frac{%
\nu }{2\tau }}a^{\dag }\right) \colon .  \label{33}
\end{eqnarray}%
Therefore we have%
\begin{eqnarray}
&&\left( -\sigma a+\mu a^{\dagger }\right) ^{m}\left \vert 0\right \rangle
\left \langle 0\right \vert \left( \tau a-\nu a^{\dagger }\right) ^{n} 
\notag \\
&=&\left( \sqrt{\frac{\sigma \mu }{2}}\right) ^{m}\left( \sqrt{\frac{\nu
\tau }{2}}\right) ^{n}\colon H_{m}\left( -\sqrt{\frac{\mu }{2\sigma }}%
a^{\dag }\right) \exp \left[ -a^{\dag }a\right] H_{n}\left( -\sqrt{\frac{%
\tau }{2\nu }}a\right) \colon ,  \label{34}
\end{eqnarray}%
where $\left \vert 0\right \rangle \left \langle 0\right \vert =\colon \exp %
\left[ -a^{\dag }a\right] \colon .$ Utilizing the following formular (for
details in Appendix) 
\begin{equation}
F\left( a^{\dagger },a\right) =%
\begin{array}{c}
\dot{\vdots}%
\end{array}%
f\left( a^{\dagger },a\right) 
\begin{array}{c}
\dot{\vdots}%
\end{array}%
=2\int \frac{d^{2}z}{\pi }%
\begin{array}{c}
\dot{\vdots}%
\end{array}%
\left \langle -z\right \vert F\left( a^{\dagger },a\right) \left \vert
z\right \rangle \exp \left[ 2\left( az^{\ast }-a^{\dag }z+a^{\dag }a\right) %
\right] 
\begin{array}{c}
\dot{\vdots}%
\end{array}%
,  \label{transformation}
\end{equation}%
we can obtain the Weyl ordering form of $\left( -\sigma a+\mu a^{\dagger
}\right) ^{m}\left \vert 0\right \rangle \left \langle 0\right \vert \left(
\tau a-\nu a^{\dagger }\right) ^{n}$, namely 
\begin{eqnarray}
&&\left( -\sigma a+\mu a^{\dagger }\right) ^{m}\left \vert 0\right \rangle
\left \langle 0\right \vert \left( \tau a-\nu a^{\dagger }\right)
^{n}=2\left( \sqrt{\frac{\sigma \mu }{2}}\right) ^{m}\left( -\sqrt{\frac{\nu
\tau }{2}}\right) ^{n}  \notag \\
&&\times 
\begin{array}{c}
\dot{\vdots}%
\end{array}%
\int \frac{d^{2}z}{\pi }H_{m}\left( \sqrt{\frac{\mu }{2\sigma }}z^{\ast
}\right) H_{n}\left( \sqrt{\frac{\tau }{2\nu }}z\right) \exp \left[ -\left
\vert z\right \vert ^{2}-2a^{\dag }z+2az^{\ast }+2a^{\dag }a\right] 
\begin{array}{c}
\dot{\vdots}%
\end{array}%
.  \label{projector}
\end{eqnarray}%
An explicit integral formula proved in Ref. \cite{fhy_2014_ijtp} 
\begin{equation}
\int \frac{d^{2}z}{\pi }H_{m}\left( z^{\ast }\right) H_{n}\left( z\right)
\exp \left[ -\left( z-\lambda \right) \left( z^{\ast }-\lambda ^{\ast
}\right) \right] =\sum_{l=0}^{\min \left[ m,n\right] }\frac{2^{2l}m!n!}{%
l!\left( m-l\right) !\left( n-l\right) !}H_{m-l}\left( \lambda ^{\ast
}\right) H_{n-l}\left( \lambda \right)  \label{twoh}
\end{equation}%
can be utilized for the derivation of Eq.(\ref{projector}). Furtherly, we
have%
\begin{equation}
\int \frac{d^{2}z}{\pi }H_{m}\left( \alpha z^{\ast }\right) H_{n}\left(
\beta z\right) \exp \left[ -\left( z-\lambda \right) \left( z^{\ast
}-\lambda ^{\ast }\right) \right] =\sum_{l=0}^{\min \left[ m,n\right] }\frac{%
\left( 4\alpha \beta \right) ^{l}m!n!}{l!\left( m-l\right) !\left(
n-l\right) !}H_{m-l}\left( \alpha \lambda ^{\ast }\right) H_{n-l}\left(
\beta \lambda \right) ,  \label{twoph}
\end{equation}%
then it follows%
\begin{eqnarray}
&&\left( -\sigma a+\mu a^{\dagger }\right) ^{m}\left \vert 0\right \rangle
\left \langle 0\right \vert \left( \tau a-\nu a^{\dagger }\right)
^{n}=2m!n!\left( -\sqrt{\frac{\sigma \mu }{2}}\right) ^{m}\left( -\sqrt{%
\frac{\nu \tau }{2}}\right) ^{n}  \notag \\
&&\times 
\begin{array}{c}
\dot{\vdots}%
\end{array}%
\sum_{l=0}^{\min \left[ m,n\right] }\frac{\left( -2\sqrt{\frac{\mu \tau }{%
\sigma \nu }}\right) ^{l}}{l!\left( m-l\right) !\left( n-l\right) !}%
H_{m-l}\left( \sqrt{\frac{2\mu }{\sigma }}a^{\dag }\right) H_{n-l}\left( 
\sqrt{\frac{2\tau }{\nu }}a\right) \exp \left[ -2a^{\dag }a\right] 
\begin{array}{c}
\dot{\vdots}%
\end{array}%
.  \label{result}
\end{eqnarray}%
Noting that Weyl ordering is invariant under the similar transformations
shown in Eq. (\ref{invariant}), and using the Weyl correspondence formula in
Eq. (\ref{rr}), we can obtain the classical correspondence of $G_{2}\left(
a^{\dag },a\right) ,$%
\begin{eqnarray}
g_{2}\left( \alpha ^{\ast },\alpha \right) &=&2\pi \text{Tr}\left[
G_{2}\left( a^{\dagger },a\right) \Delta \left( \alpha ,\alpha ^{\ast
}\right) \right] =2\pi \text{Tr}\left[ a^{\dagger m}\hat{U}\left \vert
0\right \rangle \left \langle 0\right \vert \hat{U}^{-1}a^{n}\Delta \left(
\alpha ,\alpha ^{\ast }\right) \right]  \notag \\
&=&2m!n!\left( -\sqrt{\frac{\sigma \mu }{2}}\right) ^{m}\left( -\sqrt{\frac{%
\nu \tau }{2}}\right) ^{n}\sum_{l=0}^{\min \left[ m,n\right] }\frac{\left( -2%
\sqrt{\frac{\mu \tau }{\sigma \nu }}\right) ^{l}}{l!\left( m-l\right)
!\left( n-l\right) !}H_{m-l}\left[ \sqrt{\frac{2\mu }{\sigma }}\left( \sigma
\alpha +\tau \alpha ^{\ast }\right) \right]  \label{twoh2} \\
&&\times H_{n-l}\left[ \sqrt{\frac{2\tau }{\nu }}\left( \mu \alpha +\nu
\alpha ^{\ast }\right) \right] \exp \left[ -2\left( \sigma \alpha +\tau
\alpha ^{\ast }\right) \left( \mu \alpha +\nu \alpha ^{\ast }\right) \right]
.  \notag
\end{eqnarray}

From Eq. (\ref{formula}), the classical correspondence of $G_{2}\left(
a^{\dagger },a\right) $ can also be expressed as 
\begin{eqnarray}
g_{2}\left( \alpha ^{\ast },\alpha \right) &=&e^{2\left \vert \alpha \right
\vert ^{2}}\int \frac{d^{2}\beta }{\pi }\left \langle -\beta \right \vert
a^{\dagger m}\hat{U}\left \vert 0\right \rangle \left \langle 0\right \vert 
\hat{U}^{-1}a^{n}\left \vert \beta \right \rangle e^{2\left( \alpha \beta
^{\ast }-\alpha ^{\ast }\beta \right) }  \notag \\
&=&\frac{1}{\sqrt{\mu \tau }}e^{2\left \vert \alpha \right \vert ^{2}}\int 
\frac{d^{2}\beta }{\pi }\left( -\beta ^{\ast }\right) ^{m}\beta ^{n}\exp %
\left[ -\left \vert \beta \right \vert ^{2}-2\alpha ^{\ast }\beta +2\alpha
\beta ^{\ast }-\frac{\sigma }{2\tau }\beta ^{2}-\frac{\nu }{2\mu }\beta
^{\ast }{}^{2}\right]  \label{app} \\
&=&\frac{e^{2\left \vert \alpha \right \vert ^{2}}}{\sqrt{\mu \tau }}\frac{%
\left( -1\right) ^{m+n}}{2^{m+n}}\frac{\partial ^{m}}{\partial \alpha ^{m}}%
\frac{\partial ^{n}}{\partial \alpha ^{\ast n}}\int \frac{d^{2}\beta }{\pi }%
\exp \left[ -\left \vert \beta \right \vert ^{2}-2\alpha ^{\ast }\beta
+2\alpha \beta ^{\ast }-\frac{\sigma }{2\tau }\beta ^{2}-\frac{\nu }{2\mu }%
\beta ^{\ast }{}^{2}\right] .  \notag
\end{eqnarray}%
And then using the following integral formula%
\begin{equation*}
\int \frac{d^{2}\alpha }{\pi }\exp \left[ h\left \vert \alpha \right \vert
^{2}+s\alpha +\eta \alpha ^{\ast }+f\alpha ^{2}+g\alpha ^{\ast 2}\right] =%
\frac{1}{\sqrt{h^{2}-4fg}}\exp \left[ \frac{-hs\eta +s^{2}g+\eta ^{2}f}{%
h^{2}-4fg}\right] ,
\end{equation*}%
whose convergence conditions are Re$\left( h\pm f\pm g\right) <0$ and Re$%
\left( \frac{h^{2}-4fg}{h\pm f\pm g}\right) <0,$ we have%
\begin{equation}
g_{2}\left( \alpha ^{\ast },\alpha \right) =e^{2\left \vert \alpha \right
\vert ^{2}}\frac{\left( -1\right) ^{m+n}}{2^{m+n}}\frac{\partial ^{m}}{%
\partial \alpha ^{m}}\frac{\partial ^{n}}{\partial \alpha ^{\ast n}}\exp %
\left[ -4\mu \tau \left \vert \alpha \right \vert ^{2}-2\nu \tau \alpha
^{\ast 2}-2\sigma \mu \alpha ^{2}\right] .  \label{diff}
\end{equation}%
Comparing Eqs. (\ref{twoh2}) and (\ref{diff}), we can see%
\begin{eqnarray}
&&\exp \left[ 4\mu \tau \left \vert \alpha \right \vert ^{2}+2\nu \tau
\alpha ^{\ast 2}+2\sigma \mu \alpha ^{2}\right] \frac{\partial ^{m}}{%
\partial \alpha ^{m}}\frac{\partial ^{n}}{\partial \alpha ^{\ast n}}\exp %
\left[ -4\mu \tau \left \vert \alpha \right \vert ^{2}-2\nu \tau \alpha
^{\ast 2}-2\sigma \mu \alpha ^{2}\right]  \notag \\
&=&2\left( 2\mu \sigma \right) ^{\frac{m}{2}}\left( 2\tau \nu \right) ^{%
\frac{n}{2}}\sum_{l=0}^{\min \left[ m,n\right] }\dbinom{m}{l}\dbinom{n}{l}%
l!\left( -\sqrt{\tfrac{4\mu \tau }{\sigma \nu }}\right) ^{l}  \label{nf} \\
&&\times H_{m-l}\left[ \sqrt{\tfrac{2\mu }{\sigma }}\left( \sigma \alpha
+\tau \alpha ^{\ast }\right) \right] H_{n-l}\left[ \sqrt{\tfrac{2\tau }{\nu }%
}\left( \mu \alpha +\nu \alpha ^{\ast }\right) \right] .  \notag
\end{eqnarray}%
which is a generalized differential formula related to the product of two
single-variable Hermite polynomials. If taking $\tau =\mu ^{\ast },$ $\sigma
=\nu ^{\ast }$ to meet $\left \vert \mu \right \vert ^{2}-\left \vert \nu
\right \vert ^{2}=1,\  \hat{U}$ is unitary. Thus Eq. (\ref{nf}) reduces to 
\begin{eqnarray}
&&\exp \left[ 4\left \vert \mu \right \vert ^{2}\left \vert \alpha \right
\vert ^{2}+2\mu ^{\ast }\nu \alpha ^{\ast 2}+2\mu \nu ^{\ast }\alpha ^{2}%
\right] \frac{\partial ^{m}}{\partial \alpha ^{m}}\frac{\partial ^{n}}{%
\partial \alpha ^{\ast n}}\exp \left[ -4\left \vert \mu \right \vert
^{2}\left \vert \alpha \right \vert ^{2}-2\mu ^{\ast }\nu \alpha ^{\ast
2}-2\mu \nu ^{\ast }\alpha ^{2}\right]  \notag \\
&=&2\left( 2\mu \nu ^{\ast }\right) ^{\frac{m}{2}}\left( 2\mu ^{\ast }\nu
\right) ^{\frac{n}{2}}\sum_{l=0}^{\min \left[ m,n\right] }\dbinom{m}{l}%
\dbinom{n}{l}l!\left( -\frac{2\left \vert \mu \right \vert }{\left \vert \nu
\right \vert }\right) ^{l}  \label{unitary} \\
&&\times H_{m-l}\left[ \sqrt{\tfrac{2\mu }{\nu ^{\ast }}}\left( \nu ^{\ast
}\alpha +\mu ^{\ast }\alpha ^{\ast }\right) \right] H_{n-l}\left[ \sqrt{%
\tfrac{2\mu ^{\ast }}{\nu }}\left( \mu \alpha +\nu \alpha ^{\ast }\right) %
\right] .  \notag
\end{eqnarray}%
Especially, for $m=n,$ Eq. (\ref{unitary}) is 
\begin{eqnarray}
&&\exp \left[ 4\left \vert \mu \right \vert ^{2}\left \vert \alpha \right
\vert ^{2}+2\mu ^{\ast }\nu \alpha ^{\ast 2}+2\mu \nu ^{\ast }\alpha ^{2}%
\right] \frac{\partial ^{m}}{\partial \alpha ^{m}}\frac{\partial ^{m}}{%
\partial \alpha ^{\ast m}}\exp \left[ -4\left \vert \mu \right \vert
^{2}\left \vert \alpha \right \vert ^{2}-2\mu ^{\ast }\nu \alpha ^{\ast
2}-2\mu \nu ^{\ast }\alpha ^{2}\right]  \notag \\
&=&2^{m+1}\left \vert \mu \right \vert ^{m}\left \vert \nu \right \vert
^{m}\sum_{l=0}^{m}\dbinom{m}{l}\dbinom{m}{l}l!\left( -\frac{2\left \vert \mu
\right \vert }{\left \vert \nu \right \vert }\right) ^{l}\left \vert H_{m-l}%
\left[ \sqrt{\frac{2\mu }{\nu ^{\ast }}}\left( \nu ^{\ast }\alpha +\mu
^{\ast }\alpha ^{\ast }\right) \right] \right \vert ^{2}.  \label{54}
\end{eqnarray}

\subsection{Generalized Differential Expression Related to Two-variable
Hermite Polynomials}

Following we shall derive another gereralized differential expression
related to two-variable Hermite Polynomials. From Eqs. (\ref{32}) and (\ref%
{33}) we can express the classical correspondence of $G_{2}\left( a^{\dag
},a\right) =a^{\dagger m}\hat{U}\left \vert 0\right \rangle \left \langle
0\right \vert \hat{U}^{-1}a^{n}\ $as 
\begin{eqnarray*}
g_{2}\left( \alpha ^{\ast },\alpha \right) &=&2\pi Tr\left[ \hat{U}\left(
\mu a^{\dagger }-\sigma a\right) ^{m}\left \vert 0\right \rangle \left
\langle 0\right \vert \left( \tau a-\nu a^{\dagger }\right) ^{n}\hat{U}%
^{-1}\Delta \left( \alpha ,\alpha ^{\ast }\right) \right] \\
&=&2\pi \left( -\sqrt{\frac{\sigma \mu }{2}}\right) ^{m}\left( -\sqrt{\frac{%
\nu \tau }{2}}\right) ^{n}\text{Tr}\left[ \hat{U}H_{m}\left( \sqrt{\frac{\mu 
}{2\sigma }}a^{\dag }\right) \left \vert 0\right \rangle \left \langle
0\right \vert H_{n}\left( \sqrt{\frac{\tau }{2\nu }}a\right) \hat{U}%
^{-1}\Delta \left( \alpha ,\alpha ^{\ast }\right) \right] .
\end{eqnarray*}%
Due to 
\begin{equation*}
H_{m}\left( x\right) =\sum_{k=0}^{\left[ m/2\right] }\frac{\left( -1\right)
^{k}m!}{k!\left( m-2k\right) !}\left( 2x\right) ^{m-2k},
\end{equation*}%
we can obtain%
\begin{eqnarray}
g_{2}\left( \alpha ^{\ast },\alpha \right) &=&2\pi \left( -\frac{\mu }{2}%
\right) ^{m}\left( -\frac{\tau }{2}\right) ^{n}\sum_{k=0}^{\left[ m/2\right]
}\sum_{l=0}^{\left[ n/2\right] }\frac{\left( -\frac{\sigma }{2\mu }\right)
^{k}m!}{k!\sqrt{\left( m-2k\right) !}}\frac{\left( -\frac{\nu }{2\tau }%
\right) ^{l}n!}{l!\sqrt{\left( n-2l\right) !}}  \notag \\
&&\times \text{Tr}\left[ \hat{U}\left \vert m-2k\right \rangle \left \langle
n-2l\right \vert \hat{U}^{-1}\Delta \left( \alpha ,\alpha ^{\ast }\right) %
\right] .  \label{twv}
\end{eqnarray}%
Supposed a projection operator of the number state $\left \vert
m\right
\rangle \left \langle n\right \vert \ $and from Eq. (\ref%
{transformation}), its Weyl ordering reads as 
\begin{eqnarray}
\left \vert m\right \rangle \left \langle n\right \vert &=&2\int \frac{d^{2}z%
}{\pi }%
\begin{array}{c}
\dot{\vdots}%
\end{array}%
\left \langle -z\right \vert \left. m\right \rangle \left \langle n\right.
\left \vert z\right \rangle \exp \left[ 2\left( az^{\ast }-a^{\dag
}z+a^{\dag }a\right) \right] 
\begin{array}{c}
\dot{\vdots}%
\end{array}
\notag \\
&=&2\int \frac{d^{2}z}{\pi }%
\begin{array}{c}
\dot{\vdots}%
\end{array}%
\frac{\left( -z^{\ast }\right) ^{m}z^{n}}{\sqrt{n!m!}}\exp \left[ -2\left
\vert z\right \vert ^{2}+2\left( az^{\ast }-a^{\dag }z+a^{\dag }a\right) %
\right] 
\begin{array}{c}
\dot{\vdots}%
\end{array}%
.  \label{num_weyl}
\end{eqnarray}%
By virtue of the following integral formula%
\begin{equation*}
\int \frac{d^{2}\beta }{\pi }\beta ^{\ast k}\beta ^{l}\exp \left[ -h\left
\vert \beta \right \vert ^{2}+s\beta +f\beta ^{\ast }\right] =\left(
-i\right) ^{k+l}h^{-\frac{k+l}{2}-1}e^{\tfrac{sf}{h}}H_{k,l}\left( \frac{is}{%
\sqrt{h}},\frac{if}{\sqrt{h}}\right)
\end{equation*}%
we have%
\begin{equation}
\left \vert m\right \rangle \left \langle n\right \vert =\frac{2}{\sqrt{n!m!}%
}%
\begin{array}{c}
\dot{\vdots}%
\end{array}%
H_{m,n}\left( 2a^{\dag },2a\right) \exp \left( -2a^{\dag }a\right) 
\begin{array}{c}
\dot{\vdots}%
\end{array}%
.  \label{56}
\end{equation}%
Therefore, considering Eqs. (\ref{invariant}) and (\ref{56}), we can see 
\begin{eqnarray*}
g_{2}\left( \alpha ^{\ast },\alpha \right) &=&4\pi \left( -\frac{\mu }{2}%
\right) ^{m}\left( -\frac{\tau }{2}\right) ^{n}\sum_{k=0}^{\left[ m/2\right]
}\sum_{l=0}^{\left[ n/2\right] }\frac{\left( -\frac{\sigma }{2\mu }\right)
^{k}m!}{k!\left( m-2k\right) !}\frac{\left( -\frac{\nu }{2\tau }\right)
^{l}n!}{l!\left( n-2l\right) !} \\
&&\times \text{Tr}\left[ 
\begin{array}{c}
\dot{\vdots}%
\end{array}%
H_{m-2k,n-2l}\left[ 2\left( \sigma a+\tau a^{\dagger }\right) ,2\left( \mu
a+\nu a^{\dagger }\right) \right] \exp \left[ -2\left( \sigma a+\tau
a^{\dagger }\right) \left( \mu a+\nu a^{\dagger }\right) \right] 
\begin{array}{c}
\dot{\vdots}%
\end{array}%
\Delta \left( \alpha ,\alpha ^{\ast }\right) \right] .
\end{eqnarray*}%
Weyl correspondence rule in (\ref{rr}) tells us that the classical
correspondence of $G_{2}\left( a^{\dag },a\right) $ is%
\begin{eqnarray}
g_{2}\left( \alpha ^{\ast },\alpha \right) &=&2\left( -\frac{\mu }{2}\right)
^{m}\left( -\frac{\tau }{2}\right) ^{n}\exp \left[ -2\left( \sigma \alpha
+\tau \alpha ^{\ast }\right) \left( \mu \alpha +\nu \alpha ^{\ast }\right) %
\right]  \notag \\
&&\times \sum_{k=0}^{\left[ m/2\right] }\sum_{l=0}^{\left[ n/2\right] }\frac{%
\left( -\frac{\sigma }{2\mu }\right) ^{k}m!}{k!\left( m-2k\right) !}\frac{%
\left( -\frac{\nu }{2\tau }\right) ^{l}n!}{l!\left( n-2l\right) !}%
H_{m-2k,n-2l}\left[ 2\left( \sigma \alpha +\tau \alpha ^{\ast }\right)
,2\left( \mu \alpha +\nu \alpha ^{\ast }\right) \right] .  \label{twv2}
\end{eqnarray}%
Comparing Eqs. (\ref{twv2}) and (\ref{diff}), we also derive a simplified
equation%
\begin{eqnarray}
&&\exp \left[ 4\mu \tau \left \vert \alpha \right \vert ^{2}+2\nu \tau
\alpha ^{\ast 2}+2\sigma \mu \alpha ^{2}\right] \frac{\partial ^{m}}{%
\partial \alpha ^{m}}\frac{\partial ^{n}}{\partial \alpha ^{\ast n}}\exp %
\left[ -4\mu \tau \left \vert \alpha \right \vert ^{2}-2\nu \tau \alpha
^{\ast 2}-2\sigma \mu \alpha ^{2}\right]  \notag \\
&=&2\mu ^{m}\tau ^{n}\sum_{k=0}^{\left[ m/2\right] }\sum_{l=0}^{\left[ n/2%
\right] }\frac{\left( -\frac{\sigma }{2\mu }\right) ^{k}m!}{k!\left(
m-2k\right) !}\frac{\left( -\frac{\nu }{2\tau }\right) ^{l}n!}{l!\left(
n-2l\right) !}H_{m-2k,n-2l}\left[ 2\left( \sigma \alpha +\tau \alpha ^{\ast
}\right) ,2\left( \mu \alpha +\nu \alpha ^{\ast }\right) \right] .
\label{qwe}
\end{eqnarray}%
The right of (\ref{qwe}) is a summation of two-variable Hermite Polynomials,
while Eq. (\ref{nf}) is related to the product of two single-variable
Hermite Polynomials. Eqs. (\ref{nf}) and (\ref{qwe}) are new generalized
differential expressions related to the Hermite Polynomials. For a special
case of $m=n$, Eq. (\ref{qwe}) reduces to 
\begin{eqnarray}
&&\exp \left[ 4\mu \tau \left \vert \alpha \right \vert ^{2}+2\nu \tau
\alpha ^{\ast 2}+2\sigma \mu \alpha ^{2}\right] \frac{\partial ^{m}}{%
\partial \alpha ^{m}}\frac{\partial ^{m}}{\partial \alpha ^{\ast m}}\exp %
\left[ -4\mu \tau \left \vert \alpha \right \vert ^{2}-2\nu \tau \alpha
^{\ast 2}-2\sigma \mu \alpha ^{2}\right]  \notag \\
&=&2\left( \mu \tau \right) ^{m}\sum_{k,l=0}^{\left[ m/2\right] }\frac{%
\left( -\frac{\sigma }{2\mu }\right) ^{k}m!}{k!\left( m-2k\right) !}\frac{%
\left( -\frac{\nu }{2\tau }\right) ^{l}m!}{l!\left( m-2l\right) !}%
H_{m-2k,m-2l}\left[ 2\left( \sigma \alpha +\tau \alpha ^{\ast }\right)
,2\left( \mu \alpha +\nu \alpha ^{\ast }\right) \right] .  \notag
\end{eqnarray}%
For a special case of unitary operator, e.g. $\hat{U}=\exp \left[ \frac{r}{2}%
\left( a^{\dag 2}-a^{2}\right) \right] $, we have$\  \tau =\mu ^{\ast }=\cosh
r,$ $\sigma =\nu ^{\ast }=\sinh r$, Eq. (\ref{qwe}) can be simplified to be%
\begin{eqnarray}
&&\exp \left[ 4\cosh ^{2}r\left \vert \alpha \right \vert ^{2}+2\sinh r\cosh
r\left( \alpha ^{2}+\alpha ^{\ast 2}\right) \right] \frac{\partial ^{m}}{%
\partial \alpha ^{m}}\frac{\partial ^{n}}{\partial \alpha ^{\ast n}}\exp %
\left[ -4\cosh ^{2}r\left \vert \alpha \right \vert ^{2}-2\sinh r\cosh
r\left( \alpha ^{2}+\alpha ^{\ast 2}\right) \right]  \notag \\
&=&2\cosh ^{m+n}r\sum_{k=0}^{\left[ m/2\right] }\sum_{l=0}^{\left[ n/2\right]
}\frac{\left( -\frac{\tanh r}{2}\right) ^{k+l}}{k!\left( m-2k\right) !}\frac{%
m!n!}{l!\left( n-2l\right) !}H_{m-2k,n-2l}\left[ 2\left( \alpha ^{\ast
}\cosh r+\alpha \sinh r\right) ,2\left( \alpha \cosh r+\alpha ^{\ast }\sinh
r\right) \right] .  \notag
\end{eqnarray}

\section{Applications of Eq.(\protect \ref{nf}) in the Study of Nonclassical
Features of Quantum Light Field}

For a composite system consisting of a multi-level atom (or quantum dot)
coupled to a cavity and driven by a weak coherent field, quantum-optical
effects can be demonstrated in the interaction processes of photon emission
and absorbtion with atom between ground and excited states \cite%
{Rempe_1987_prl,Schuster_2007_nature,Jundt_2008_prl,Huang_2014_pra}. In Ref. 
\cite{Majumdar_2012_pra}, author has considered a weak coherent incident
field, and the quantum state of emitted light can be expressed as a series
of excited\ coherent states $\sum_{m}C_{m}a^{\dag m}\left \vert \alpha
\right \rangle $ or a superposition of the different Fock states $%
\left
\vert \psi \right \rangle =\sum_{n}C_{n}\left \vert n\right \rangle $
(due to $\left \vert \alpha \right \rangle =\sum_{n}\frac{\alpha ^{n}}{\sqrt{%
n!}}\left \vert n\right \rangle $). Here we consider a strong coupling case,
and\ the initial state of the incident source is a single-mode squeezed
vacuum field. Only considering an effective measurement to optical field
(e.g. photon-statistics methods), this multi-photon processes originated
from quantum nonlinearity can be monitored by%
\begin{equation}
\rho \left( r,n\right) =C_{n}^{-1}a^{\dag }{}^{n}S\left( r\right) \left
\vert 0\right \rangle \left \langle 0\right \vert S^{-1}\left( r\right)
a^{n}.  \label{aa2}
\end{equation}%
which can exhibit similar behavior to that of an excited quantum state (e.g. 
$a^{\dag m}\left \vert \varphi \right \rangle $). $C_{n}=Tr\left[ \rho
\left( r,n\right) \right] =n!\cosh ^{n}rP_{n}\left( \cosh r\right) $ is a
normalized constant. $P_{n}\left( \cosh r\right) $\ is the expression of the
Legendre polynomials. $S\left( r\right) =$ $\exp \left[ \frac{r}{2}\left(
a^{\dag 2}-a^{2}\right) \right] $ denotes a unitary operator with the
following transformation identities%
\begin{eqnarray}
S^{-1}\left( r\right) aS\left( r\right) &=&a\cosh r+a^{\dag }\sinh r,  \notag
\\
S^{-1}\left( r\right) a^{\dag }S\left( r\right) &=&a^{\dag }\cosh r+a\sinh r.
\label{aa1}
\end{eqnarray}%
In order to investigate the non-classical features of $\rho \left(
r,n\right) ,$ Eq. (\ref{ww}) shows that its quasi-probability distribution
Wigner function can be written as%
\begin{eqnarray}
W\left( \alpha ,\alpha ^{\ast }\right) &=&2\pi \text{Tr}\left[ \rho \left(
r,n\right) \Delta \left( \alpha ,\alpha ^{\ast }\right) \right]  \notag \\
&=&2\pi C_{n}^{-1}\text{Tr}\left[ S\left( r\right) \left( a^{\dag }\cosh
r+a\sinh r\right) ^{n}\left \vert 0\right \rangle \left \langle 0\right
\vert \left( a\cosh r+a^{\dag }\sinh r\right) ^{n}S^{-1}\left( r\right)
\Delta \left( \alpha ,\alpha ^{\ast }\right) \right] .  \label{aa3}
\end{eqnarray}%
Comparing (\ref{aa1}) with (\ref{invers}), we can see $\mu =\tau \rightarrow
\cosh r,$ $\sigma =\nu \rightarrow -\sinh r.$ From Eqs. (\ref{aa3}) and (\ref%
{weylc}), thus we can obtain the Wigner function of $\rho \left( r,n\right)
, $%
\begin{eqnarray*}
W\left( \alpha ,\alpha ^{\ast }\right) &=&\frac{1}{P_{n}\left( \cosh
r\right) }\left( \frac{-\sinh r}{2}\right) ^{n}\exp \left[ -2\left \vert
\alpha \sinh r-\alpha ^{\ast }\cosh r\right \vert ^{2}\right] \\
&&\times \sum_{l=0}^{n}\left( 
\begin{array}{c}
n \\ 
l%
\end{array}%
\right) \frac{2^{l}\left( -\coth r\right) ^{l}}{\left( n-l\right) !}\left
\vert H_{n-l}\left[ i\sqrt{\frac{2}{\tanh r}}\left( \alpha ^{\ast }\cosh
r-\alpha \sinh r\right) \right] \right \vert ^{2}.
\end{eqnarray*}

\begin{figure}[h]
\centering
\includegraphics[scale=0.8]{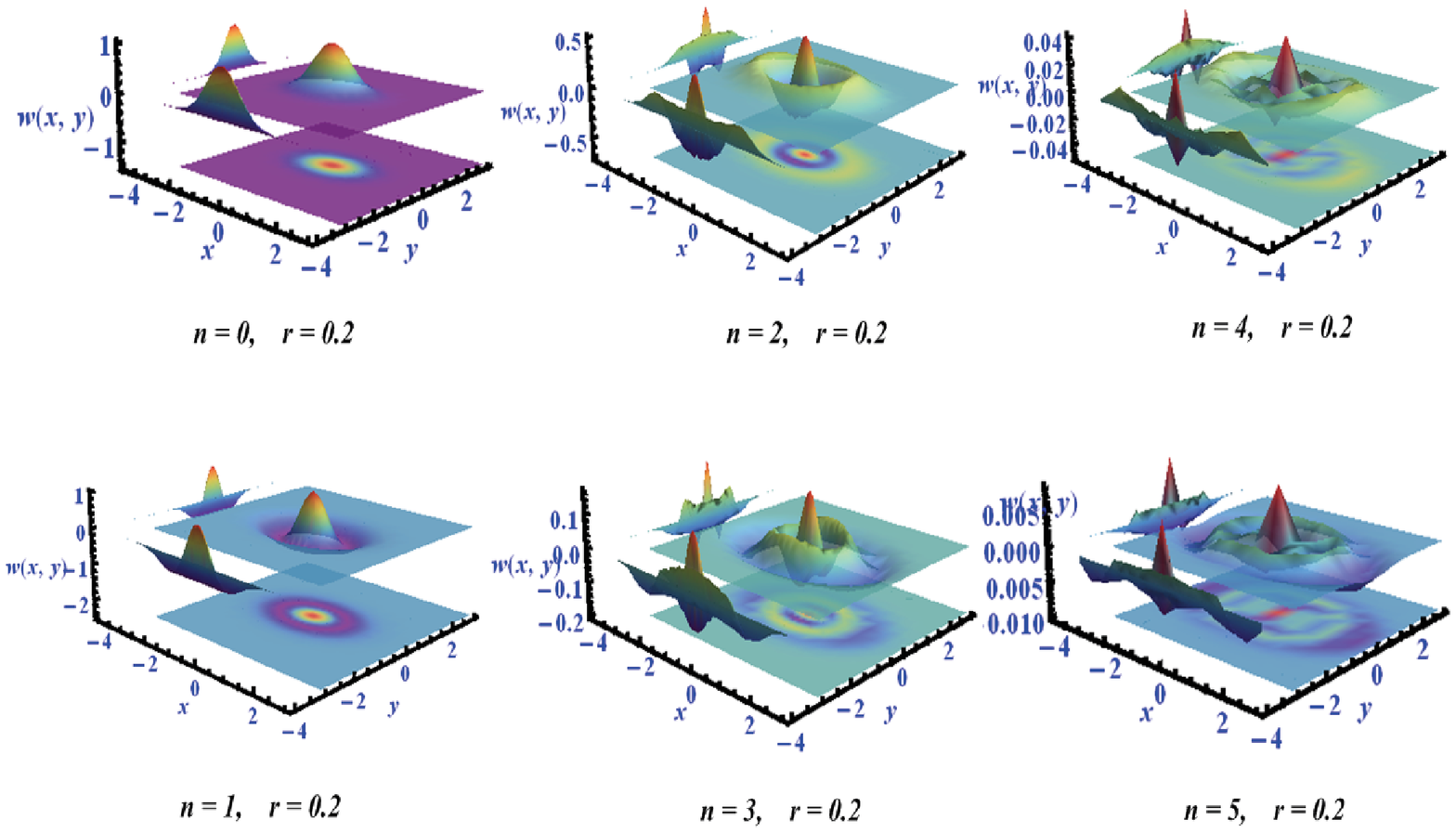}
\caption{\textsf{Wigner distributions of the excited squeezed vacuum state
with fixed squeezing. In the top row, even number photons have been added to
the squeezing vacuum field. The case of odd number photons has been shown in
the bottom row.}}
\end{figure}
In Fig. 1, negative part in certain region of phase-space indicates an
evidence of nonclassicality of the state generated by adding photon to a
weak squeezed radiation field. The variance of photon-addition can exhibit
different nonclassical features with the fixed squeezing value of $r=0.2$,
which dominates the quadrature distribution in the directions of $-X$ and $%
-Y $. The top row with three pictures shows that even photon have been added
to the weak squeezed radiation field, odd photon at bottom row.

\section{Acknowledgements}

This work has been supported in part by the Natural Science Foundation of
China (NSFC, No.61205115 and No.11204004), Outstanding Young Talent
Foundation of Anhui Province Colleges and Universities No.2012SQRL040, and
also Natural Science Foundation of Anhui Province Colleges \& Universities
under Grant KJ2012Z035. The authors acknowledge fruitful discussions about
physics meaning with Professor Fan Hong-yi.

\section{Appendix: Derivation of Eq. (\protect \ref{transformation})}

The Glauber-Sudarshan P representation, as one of most important quantum
phase space theory, is the quasiprobability distribution in which
observables are expressed in normal order. By virtue of the coherent state
representation $\left \vert z\right \rangle =\exp \left( za^{\dag }-z^{\ast
}a\right) \left \vert 0\right \rangle ,$ the $P$-representation of a quantum
density matrix $\rho $ is defined as%
\begin{equation}
\rho =\int \frac{d^{2}z}{\pi }P\left( z,z^{\ast }\right) \left \vert z\right
\rangle \left \langle z\right \vert .  \label{a1}
\end{equation}%
The inverse relation of Eq. (\ref{a1}) is%
\begin{equation}
P\left( z,z^{\ast }\right) =e^{\left \vert z\right \vert ^{2}}\int \frac{%
d^{2}\beta }{\pi }\left \langle -\beta \right \vert \rho \left \vert \beta
\right \rangle \exp \left( \left \vert \beta \right \vert ^{2}+z\beta ^{\ast
}-z^{\ast }\beta \right) ,  \label{Mehta}
\end{equation}%
which was first obtained by Mehta \cite{Mehta_prl_1967}. Substituting (\ref%
{Mehta}) into (\ref{a1}) and Utilizing Weyl ordering of the coherent state
projector, namely%
\begin{equation*}
\left \vert z\right \rangle \left \langle z\right \vert =2%
\begin{array}{c}
\dot{\vdots}%
\end{array}%
\exp \left[ -2\left( z^{\ast }-a^{\dag }\right) \left( z-a\right) \right] 
\begin{array}{c}
\dot{\vdots}%
\end{array}%
,
\end{equation*}%
we have%
\begin{eqnarray}
\rho &=&2\int \frac{d^{2}z}{\pi }e^{\left \vert z\right \vert ^{2}}\int 
\frac{d^{2}\beta }{\pi }\left \langle -\beta \right \vert \rho \left \vert
\beta \right \rangle \exp \left( \left \vert \beta \right \vert ^{2}+z\beta
^{\ast }-z^{\ast }\beta \right) 
\begin{array}{c}
\dot{\vdots}%
\end{array}%
\exp \left[ -2\left( z^{\ast }-a^{\dag }\right) \left( z-a\right) \right] 
\begin{array}{c}
\dot{\vdots}%
\end{array}
\notag \\
&=&2\int \frac{d^{2}\beta }{\pi }%
\begin{array}{c}
\dot{\vdots}%
\end{array}%
\left \langle -\beta \right \vert \rho \left \vert \beta \right \rangle \exp %
\left[ 2\left( a\beta ^{\ast }-a^{\dag }\beta +a^{\dag }a\right) \right] 
\begin{array}{c}
\dot{\vdots}%
\end{array}%
.  \label{a2}
\end{eqnarray}%
which can conveniently recast operators into their Weyl ordering.


\begin{thebibliography}{99}
\bibitem{Hochstadt_ie_1973} See e.g. H. Hochstadt, \emph{Integral Equations,}
Wiley (1973) New York.

\bibitem{fhy_jopb_2003} H. Y. Fan, J. Opt. B: Quantum Semiclass. Opt. 
\textbf{5} (2003) R1--R17.

\bibitem{fhy_ctp_2008} H. Y. Fan and X. B. Tang, Commun. Theor. Phys. 
\textbf{49} (2008) 1169--1172.

\bibitem{fhy_ap_2006} H. Y. Fan, H. L. Lu and Y. Fan, Ann. Phys. \textbf{321}
(2006) 480

\bibitem{erdelyi_1953_bmp} A. Erdelyi, \emph{Higher Transcendental
Functions: The Batemann Manuscript Project}, McGraw-Hill (1953) New York.

\bibitem{Namias_jimia_1980} V. Namias, J. Inst. Math. Its Appl. \textbf{25}
(1980) 241.

\bibitem{Mendlovic_josaa_1993} D. Mendlovic and H. M. Ozaktas, J. Opt. Soc.
Am. \textbf{A10} (1993) 1875.

\bibitem{Bernado_oc_1994} Luis M. Bernado and Oliverio D. D. Soares, Opt.
Commun. \textbf{110} (1994) 517.

\bibitem{fhy_ol_2004} H. Y. Fan and X. F. Xu, Opt. Lett. \textbf{29} (2004)
1048-1050.

\bibitem{Weyl_zp_1927} H. Weyl, Z. Phys. \textbf{46} (1927) 1.

\bibitem{Weyl_tgqm_2007} H. Weyl (Author), H. P. Robertson (Translator), 
\emph{The Theory of Groups and Quantum Mechanics,} Kessinger Publishing
(2007) LLC.

\bibitem{Wigner_pr_1932} E. P. Wigner, Phys. Rev. \textbf{40} (1932) 749

\bibitem{fhy_jpa_1992} H. Y. Fan, Phys. A: Math. Gen. \textbf{25} (1992)
3443-3447.

\bibitem{fhy_pla_2002} H. Y. Fan and J. H. Chen Phys. Lett. \textbf{A303}
(2002) 311.

\bibitem{fhy_ctp_2003} H. Y. Fan, Commun. Theor. Phys. \textbf{40} (2003)
409--414.

\bibitem{chen_ijtp_2012} J. H. Chen, H. Y. Fan and X. B. Tang, Int. J.
Theor. Phys. \textbf{51} (2012) 14.

\bibitem{fhy_2014_ijtp} H. Y. Fan and Z. Wang, Int. J. Theor. Phys. \textbf{%
53} (2014) 964--970.

\bibitem{fhy_1991_jpa} H. Y. Fan and J. Vanderlinde, J. Phys. \textbf{A24}
(1991) 2529.

\bibitem{Glauber_1963_pr} R. J. Glauber, Phys. Rev. \textbf{130} (1963)
2529; \textbf{131} (1963) 2766.

\bibitem{Rempe_1987_prl} G. Rempe, H. Walther, and N. Klein, Phys. Rev.
Lett. \textbf{58} (1987) 353.

\bibitem{Schuster_2007_nature} D. I. Schuster, A. A. Houck, J. A. Schreier,
A. Wallraff, J. M. Gambetta, A. Blais, L. Frunzio, J. Majer, B. Johnson, M.
H. Devoret, S. M. Girvin, and R. J. Schoelkopf, Nature \textbf{445} (2007)
515.

\bibitem{Jundt_2008_prl} G. jundt, L. Robledo, A. H\"{o}gele, S. F\"{a}lt
and A. Imamo\v{g}lu, Phys. Rev. Lett. \textbf{100} (2008) 177401.

\bibitem{Huang_2014_pra} J. F. Huang and C. K. Law, Phys. Rev. \textbf{A89}
(2014) 033827.

\bibitem{Majumdar_2012_pra} A. Majumdar, M. Bajcsy, and J. Vu\v{c}kovi\'{c},
Phys. Rev. \textbf{A85} (2012) 041801(R).

\bibitem{Mehta_prl_1967} C. L. Mehta, Phys. Rev. Lett. \textbf{18} (1967) 752
\end{thebibliography}
\end{document}